\documentclass[letterpaper,10pt]{article} 

\usepackage{opticameet3} 

\newcommand\authormark[1]{\textsuperscript{#1}}

\usepackage{amsmath,amssymb}
\usepackage{hyperref} 
\usepackage[
backend=biber,
style=numeric,
sorting=ynt,
hyperref=true
]{biblatex}
\usepackage{subcaption}
\begin{document}

\title{Linking Fluorescence Spectroscopy and 31P MRS in NADH Phantoms}

\author{Giorgi Asatiani\authormark{1,*}, Arthur Gautheron,\authormark{1,2} , Laurent Mahieu-Williame\authormark{1}, Hélène Ratiney,\authormark{1}  and Bruno Montcel\authormark{1,**}}

\address{

\authormark{1}INSA‐Lyon, Universite Claude Bernard Lyon 1, CNRS, Inserm, CREATIS UMR 5220, U1294, F‐69621, LYON, France \\
\authormark{2} CPE, Lyon, France \\
}

\email{\authormark{*}giorgi.asatiani@creatis.insa-lyon.fr\\
\authormark{**}bruno.montcel@univ-lyon1.fr} 

\begin{abstract}
Glioblastoma exhibits significant metabolic alterations, making tracking energy metabolism important for its characterization.
This could be crucial for glioblastoma resection in neurosurgery.
We link two NADH monitoring methods, showing linear dependence on phantom concentrations. 
\end{abstract}

\section{Introduction}
One of the most important challenges that results in high recurrence and mortality rates in Glioblastoma patients, alongside the low cure rate, is boundary detection.
As the margin of the tumor is frequently very complicated to differentiate from the healthy tissue it becomes hard to resect and completely cut it out during neurosurgery. 

The tumor detection methods include preoperative approaches, such as Magnetic Resonance Spectroscopy (MRS), which relies on analyzing metabolic activity by detecting biochemical markers.
The second way is intraoperative technique like fluorescence spectroscopy, which depends on the use of fluorescent agents or endogenous fluorescent molecules to highlight tumor tissue through fluorescence.

Nicotinamide adenine dinucleotide (NAD) is a crucial coenzyme involved in energy metabolism and redox reactions in all living cells. It exists in two forms, NAD+ (oxidized) and NADH (reduced), which interconvert during metabolic processes like glycolysis, the tricarboxylic acid cycle, and the electron transport chain. Beyond energy metabolism, NAD+ and NADH are implicated in various biological activities, including calcium homeostasis, antioxidation, gene expression, immune function, aging, and cell death. The intracellular redox state is largely regulated by the NAD+/NADH ratio, which is a key indicator of cellular redox balance. Changes in this ratio are associated with various physiological and pathological conditions, such as aging, diabetes, stroke, cancer, and epilepsy.

Our goal is to link these two ways of measurement for improvement of the boundary detection of the tumor that would help neurosurgeon in operation room.
Therefore we will be looking at the NADH phantoms, as this metabolite is suggested to be a good marker in tumor detection. Multiple groups have investigated it as a parameter and have come to such conclusion  \cite{Sun-et-al-2010} \cite{butte-et-al} \cite{Liu-et-al-2018} and it's a good choice as it is both fluorescent and can be measured through $^{31}P$ 
MRS\cite{IntraCellularRedox}.

\section{Materials and Methodology}
Throughout experiments, we have been using $4$ distinct phantoms of NADH solution diluted in deionized water with exact concentrations of $1mM$, $5mM$, $10mM$, and $15mM$. These solutions are put inside a plastic, $15ml$, transparent, falcon tube with a conic bottom.
\subsection{Optical setup}

For the fluorescence measurements, the setup, described in \cite{Gautheron2023Jun}, consists of a probe including excitation and collection fibers (Collection fiber with the NA=0.22) that is connected to the spectrometer through a high-pass filter with a cutoff wavelength of 430 nm, and also to an power controlled excitation laser.
We place the probe perpendicularly to the side of the tube and use the excitation wavelength of $405nm$ as it can be used for not only the excitation of NADH but also the flavins (FAD), the lipopigments, the porphyrins and the chlorins, which can be useful for further, in-vivo investigations.

We collected the spectrum $10$ times with $200ms$ integration time and averaged them out. After that, we recorded the signature from just water inside the tube to have the measurement of the background and we subtracted the average values of the spectrum of water from the spectrum of the phantoms.
\subsection{MRS setup}

We performed MRS experiments using a $1H/31P$ volume coil  (40mm diameter) on a 11.7T Brucker MR imaging system. Acquisitions were done using a simple non-localized sequence for each tube: 31P single pulse acquisition with  $90^\circ$ block pulse ($0.16 \ ms$ duration), detection bandwidth of $8000$ Hz;  with 64 averages, $2s$ repetition time. Total acquisition time was $2min \ 8s$. We repeated this single pulse acquisition 3 times and averaged out the results for further calculations, including the calculations of standard deviation.
\section{Results}
\begin{figure}[ht]
    \centering
    \begin{subfigure}{0.47\textwidth}
        \includegraphics[width=\linewidth]{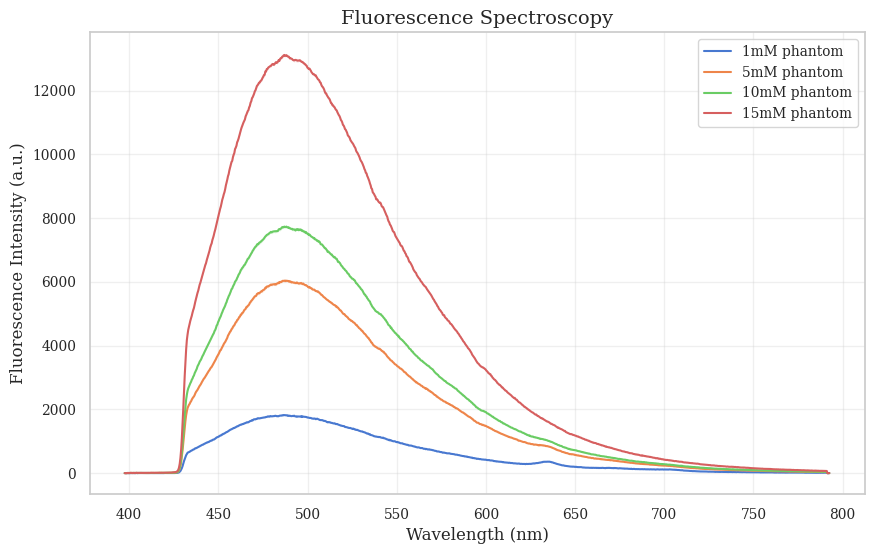}
                \caption{}
    \end{subfigure}
    \hfill
    \begin{subfigure}{0.47\textwidth}
        \includegraphics[width=\linewidth]{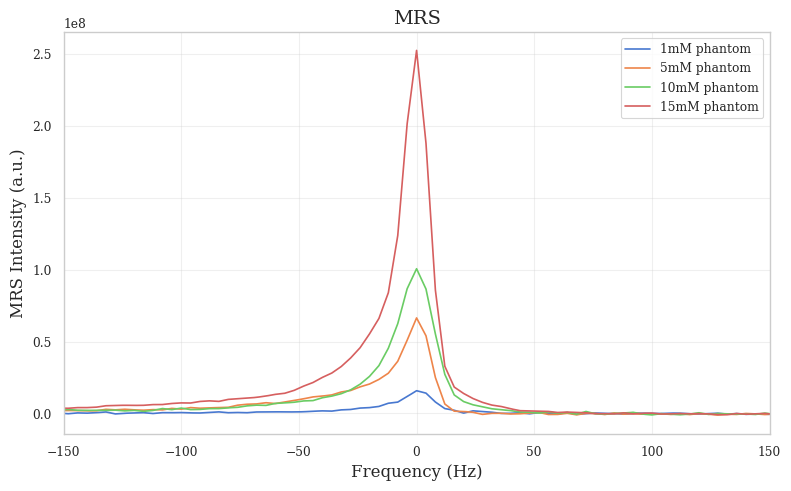}
        \caption{}
    \end{subfigure}    
    \caption{\textit{(a) Fluorescence spectroscopy. Data was obtained by averaging out 10 acquisitions using $405nm$ excitation light. Probe was placed perpendicularly to the side of the transparent tube with NADH solution. Total integration time was $2s$.
    (b) MRS. The results were obtained by using 31P single block pulse acquisition with 64 averages, $2s$ repetition time ($2min \ 8s$ total acquisition time). This sequence was used $3$ times and the mean of those three spectra was calculated}}
    \label{fig:Phantom_Aqcuisitions}
\end{figure}

Our obtained results describing the different concentration acquisitions for both fluorescence and MRS can be seen in Figure \ref{fig:Phantom_Aqcuisitions}.
 \\
MRS data was acquired in the frequency domain with the peak centered at 0 Hz. We extracted the real part of the spectrum, and applied phase correction.

For both of these methods of measurements the Signal to Noise Ratio was calculated. In case of Fluorescence the noise was taken as standard deviation of the spectrum recorded with only water inside the tube, and in case of MRS in each acquisition it's defined as peak over the standard deviation of noise that can be seen outside the region of the peak in frequency domain.

The results of Signal to Noise Ratios (SNR) depending on concentration can be seen in Table \ref{tab:SNRs}

\begin{table}[]
    \centering
    \begin{tabular}{|c|c|c|}
     \hline
     Concentration (mM) & Fluorescence spectroscopy SNR & MRS SNR\\
      \hline
     1    & 18.4 $\pm$ 0.2 & 28.3 $\pm$ 2.6 \\
     \hline
     5    & 60.8 $\pm$ 0.3 & 118.2 $\pm$ 3.8 \\
     \hline
     10   & 77.7 $\pm$ 0.3 & 175.1 $\pm$ 10.0 \\
     \hline
     15   & 131.6 $\pm$ 0.4 & 496.2 $\pm$ 69.9 \\
     \hline
    \end{tabular}
    \caption{\textit{SNR dependence on concentration for fluorescence spectroscopy and for MRS. Values represent mean $\pm$ standard deviation.}}
    \label{tab:SNRs}
\end{table}

\begin{figure}[ht]
    \centering
    \includegraphics[width=0.6\linewidth]{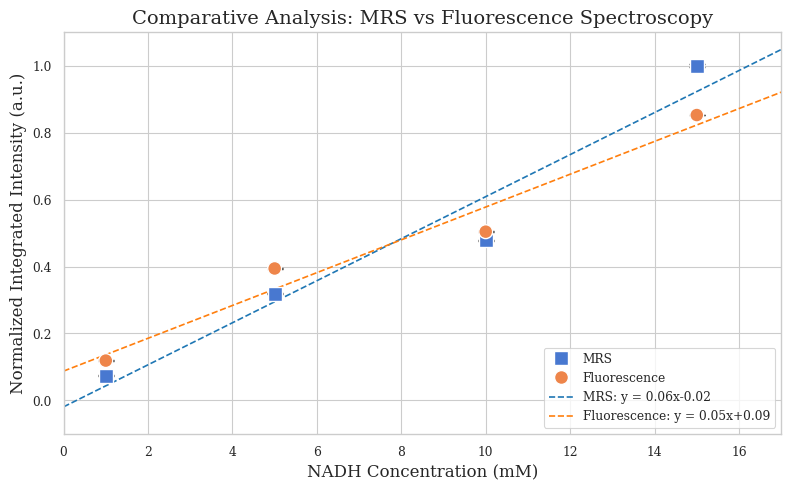}
    \caption{\textit{We performed a linear regression analysis comparing fluorescence spectroscopy and MRS results. The analysis used peak integrals from both methods across different concentrations.
    First, MRS values were scaled by dividing them by a scaling factor. This factor was the ratio of the total MRS results to the total fluorescence spectroscopy results.
    After scaling, both datasets were normalized. Each dataset was divided by the highest value across both methods, setting the maximum value to 1.}}
    \label{fig:Int_vs_conc}
    \vspace*{-0.5cm}
\end{figure}

From these results, the integrals of each peak have been calculated as the measure of the NADH inside the probed volume.
The dependence of the measured integral on the values of the concentration are of course expected to be linear and both MRS and Fluorescence Spectroscopy follow this trend. 
For a possible comparison between two methods we scaled the obtained values of integrals from MRS by the scaling factor that was the ratio in between the sum of all data points in Fluorescence spectroscopy and MRS. Linear regression fit was performed in both cases and final results can be seen in Figure \ref{fig:Int_vs_conc} 

\section{Discussion and Conclusions}
We suggest a possibility of linkage in between the two measurement methods. Studies carried out on phantoms establishes the linear correlation between MRS and fluorescence spectroscopy. MRS results suggest significantly higher SNR, especially for higher concentrations but that is expected as the volume that is investigated by single non-localized sequence is significantly higher then the probing volume for fluorescence spectroscopy. 

Further investigations would include the simultaneous acquisitions using localized sequence to have the same probing volume with same acquisition time. 

\section{Acknowledgments}
This work was performed within the framework of the LABEX PRIMES (ANR-11-LABX-0063) of Université de Lyon, within the program "Investissements d'Avenir" (ANR-11-IDEX-0007) operated by the French National Research Agency (ANR). This material is based upon work done on the ISO 9001:2015 PILoT facility , a facility of France Life Imaging network (grant ANR-11-INBS-0006).


\end{document}